\documentclass[doublecol]{epl2} 
\usepackage{epsfig}
\usepackage{graphicx}
\usepackage{bm}
\usepackage{amssymb}
\usepackage{amsmath}
\usepackage{epstopdf}

\begin{document}

\title{Efimov States of Heavy Impurities in a Bose-Einstein Condensate}
\shorttitle{Efimov States of Heavy Impurities in Bose-Einstein Condensate}

\author{N.~T. Zinner}
\institute{Department of Physics and Astronomy, Aarhus University, Aarhus C, DK-8000}

\date{\today}

\abstract{
We consider the problem of two heavy impurity particles embedded in a gas of weakly-interacting 
light mass bosonic particles in the condensed state. Using the Bogoliubov approach to describe the 
bosonic gas and the Born-Oppenheimer approximation for the three-body dynamics, we calculate the
modification to the heavy-heavy two-body potential due to the presence of the condensate. For the 
case of resonant interaction between the light bosons and the impurities, we
present (semi)-analytical results for the potential in the limit of a large condensate 
coherence length. In particular, we find a formula for the modification of the Efimov scaling 
factor due to the presence of a degenerate bosonic gas background.}

\maketitle

\section{Introduction}
The three-body spectrum for two heavy and one light particle with short-range resonant interactions between 
the heavy and light particles can be treated using the Born-Oppenheimer approximation. One 
decouples the slow degrees of freedom (relative motion of the two heavy particles) from the 
motion of the light particle to obtain an effective heavy-heavy potential. What is found
is the famous Efimov spectrum of geometrically scaling three-body states where subsequent 
states differ by a simple factor of the form $e^{\pi/s_0}$ where $s_0$ is called the scaling parameter
\cite{efimov1970}. 
For equal mass particles, $e^{\pi/s_0}\sim 22.7$, which is a severe limiting factor in the 
quest for observing many states. However, in the heavy-heavy-light system the factor 
can be much smaller \cite{jensen2004,braaten2006}. This has generated a lot of interest in studying
three-body systems with mixed atomic species in the ultracold regime \cite{ferlaino2010}.

The ultracold gases used for three-body experiments are typically at very low densities.
The three-body dynamics can thus be described by assuming that three-body states are essentially 
isolated entities in spite of the fact that there is always some finite density background present.
However, experiments should be able to push into higher densities and thus be able to investigate
the effects of the finite background on few-body properties. Some recent works have been 
devoted to the question of what happens when the particles are fermions that have to obey the 
Pauli principle \cite{nishida2009,macneill2011,nygaard2011,song2011,niemann2012}. Here we will address 
the related issue of what happens when there is a Bose gas background that is in the condensate
phase. A few other papers have recently discussed this type of question from the many-body
point-of-view \cite{borzov2012,zhou2012}. Here we take a different point-of-view and consider the
background condensate as a perturbation on the three-body state. For simplicity we will
consider the case of two heavy impurity particles in a Bose gas of light bosons. For the
two-body case the problem of an impurity in a condensate with strong interactions was
revived recently \cite{cucchietti2006,kalas2006,tempere2009}, and likewise for the 
case of multiple such impurities\cite{casteels2011,santamore2011}. The Bose-Fermi mixture
in cold atomic gases was discussed in detail about a decade ago \cite{bijlsma2000,viverit2000}
and the later works address a particularly interesting regime with large imbalance in the 
populations. However, we are not aware of any works discussing the presence of an Efimov
effect in a condensate background and the influence that the background can have on the
universal three-body spectrum.

Our approach will be based on the Bogoliubov approximation for the condensate dynamics. We will 
work in the limit of a weakly-coupled Bose gas with repulsive short-range interactions, i.e. small and positive
scattering length, $a_{B}>0$. The impurity-light boson interaction will be allowed to take on 
any value. The condensate coherence length, $\xi=1/\sqrt{8\pi n_0 a_B}$ with $n_0$ the condensate gas density, 
will therefore be large and its inverse a useful expansion parameter. The problem we solve is 
very similar to the classical problem of electron-electron interaction mediated by phonons; 
the impurities are the (heavy) electrons while the light bosons are the phonons. Of course, 
in the case of a Bose condensate of massive particle, the phonon dispersion is 
only linear for low momenta and eventually becomes quadratic. This will be properly taken 
into account in our framework. The problem considered here is closely related to the
bosonic Kondo problem  with two impurities and a bosonic 
bath \cite{affleck1995,smith1999,sengupta2000,vojta2000,zarand2002,florens2006}. The study
here should thus be of interest for both cold atomic gases and for condensed-matter 
physics.

\section{Theoretical Model}\label{theory}
We consider two heavy impurities of mass $M$ and a gas of light bosonic particles of mass 
$m\ll M$. The Hamiltonian is
\begin{eqnarray}
&H=\sum_{\bm k} \epsilon_I(\bm k)c_{\bm k}^{\dagger}c_{\bm k}+\sum_{\bm k}\epsilon_B(\bm k)b_{\bm k}^{\dagger}b_{\bm k}&\nonumber\\
&+U_{B}\sum_{\bm q}n_B({\bm q})n_B({-\bm q})+U_{IB}\sum_{\bm q}n_B({\bm q})n_I({-\bm q}),&
\end{eqnarray}
where $c$ is the impurity and $b$ the boson operator.
We use zero-range density-density 
impurity-boson and boson-boson interactions with $n_I(\bm q)=\sum_{\bm k}c_{\bm k+\bm q}^{\dagger}c_{\bm k}$ and
$n_B(\bm q)=\sum_{\bm k}b_{\bm k+\bm q}^{\dagger}b_{\bm k}$. 
The dispersions are $\epsilon_I(\bm k)=\hbar^2{\bm k}^2/2M$ and $\epsilon_B(\bm k)=\hbar^2{\bm k}^2/2m$. In the 
weakly-coupled limit we have $U_B=4\pi\hbar^2a_B/m$. The parameter $U_{IB}$ will be discussed later.
We use Bogoliubov theory to 
describe the light bosonic particles \cite{fetter1971}. 
We therefore transform to quasi-particles, $\gamma_{\bf k}$
and $\gamma_{\bm k}^\dagger$, in the standard way. Furthermore, we will assume that the 
condensate density is small so that the number of quasi-particles is also small. This 
allows us to neglect all terms except $\gamma_{\bm k}^{\dagger}\gamma_{\bm k}$ in the
transformed Hamiltonian. Dropping unimportant constant terms, the bosonic dispersion 
and the interaction term becomes
\begin{eqnarray}
\sum_{\bm k\neq 0}E(\bm k)\gamma_{\bm k}^{\dagger}\gamma_{\bm k}
+U_{IB}\sum_{\bm q\bm k\bm k'}c_{\bm k-\bm q}^{\dagger}c_{\bm k}\gamma^{\dagger}_{\bm k'+\bm q}\gamma_{\bm k'},
\end{eqnarray}
where $E(\bm k)=\sqrt{U_B n_0{\hbar^2\bm k}^2/m_B+({\hbar^2\bm k}^2/2m_B)^2}$.
This corresponds to impurity particles with dispersion $\epsilon_I(\bm k)$ interacting with Bose gas particles with 
dispersion $E(\bm k)$ through a contact interaction with strength $U_{IB}$. In the case where $E(\bm k)$ is 
linear in $\bm k$, this corresponds to a system of (heavy) electrons interacting with phonons through a non-dispersive
zero-range intearction.

We now proceed to solve the three-body problem of 
two heavy impurities and one light bosonic quasi-particle.
Note that these states should be consider resonances similar
to the three-boson case in recent experiments \cite{ferlaino2010}.
The absolute ground state should be a bound state containing 
both impurities and all the bosons which is not experimentally 
realized in these dilute atomic gases. We also note that there
can be two light bosons and one impurity three-body bound states 
in the system. However, this configuration of masses strongly 
disfavours the Efimov effect ($e^{\pi/s_0}$ is very large \cite{jensen2004,braaten2006}) 
and will not be discussed here.

\subsection{Born-Oppenheimer Approximation} 
As we have just argued, we can model the interaction of the impurities and the 
bosons via
a zero-range interaction, and we thus write 
$V(\bm r)=U_{IB}\left[\delta(\bm r-\bm R/2)+\delta(\bm r+\bm R/2)\right]$,
where we assume that the two heavy impurities are located at $\pm \bm R/2$. This potential needs to be 
regularized since as it stands it leads to an ultraviolet divergence. We return to this point below.
The essence of the Born-Oppenheimer approximation is that we first solve for the dynamics 
of the light bosonic particles while assuming that $\bm R$ is fixed, and then proceed to consider
the Schr{\"o}dinger equation for the two heavy particles as function of $\bm R$. The 
relative distance of the heavy particles, $\bm R$, is our adiabatic variable which we assume
changes on a much slower timescale than the positions of the bosonic particles.

Now we consider the Schr{\"o}dinger
equation for the particle of mass $m$ in this potential
%\begin{equation}
$H\phi=E_R\phi$,
%\end{equation}
where $E_R$ is the energy and $\phi$ the wave function of a light bosonic (quasi)-particle. 
Since $V(\bm r)$ contains delta-functions, it is 
convenient to work in momentum-space. 
The Schr{\"o}dinger equation in momentum-space can then be written
\begin{equation}
E({\bm k})\phi(\bm k)+\frac{1}{(2\pi)^3}\int d^3k' \phi(\bm k')V(\bm k-\bm k')=E_R\phi(\bm k),
\end{equation}
where $V(\bm q)=2U_{IB} \cos\left(\bm q\cdot \frac{\bm R}{2}\right)$.
Integration over ${\bm k}$ on both sides, assuming that $\phi$ is an even function of ${\bm k}$ ($s$-wave solutions), and elementary 
trigonometric manipulations reduce the equation to the form
\begin{equation}
1=-\frac{U_{IB}}{(2\pi)^3}\int d^3k \frac{1}{E({\bm k})-E_R}-\frac{U_{IB}}{(2\pi)^3}\int d^3k \frac{\cos\left(\bm k\cdot\bm R\right)}{E({\bm k})-E_R}.
\end{equation}

We now relate $U_{IB}$ and the scattering length of the interaction between the impurity and the 
bosonic particles, $a$. While this can be done most elegantly by using Tan's pseudopotential \cite{tan2008,valiente2012},
we use a traditional approach that is very easy in the Born-Oppenheimer limit.
From the Lippmann-Schwinger equation for the impurity-boson scattering we have
\begin{equation}
\frac{1}{U_{IB}}=\frac{\mu}{2\pi a\hbar^2}-\frac{1}{(2\pi)^3}\int d^3k\frac{1}{\epsilon^{\mu}_{\bm k}},
\end{equation}
where $\mu=mM/(m+M)$ and $\epsilon^{\mu}_{\bm k}=\hbar^2k^2/2\mu$ are reduced mass and energy. Since we assume 
$m\ll M$, we can safely use $\mu=m$ and $\epsilon^{\mu}_{\bm k}=\epsilon_{\bm k}$. Note that we do use the 
bare dispersion of the bosons, $\hbar^2{\bm k}^2/2m$, and {\it not} $E(\bm k)$. This is necessary since the 
heavy-light scattering length, $a$, is defined in vacuum and the Lippmann-Schwinger problem must therefore also be 
solved in vacuum.
Inserting the relation between $U_{IB}$ and $a$ we arrive at our central equation
\begin{align}
&\frac{R}{a}=-\frac{2}{\pi}\alpha R \int_{0}^{\infty} dx \left[\frac{x^2}{[x^4+A^2x^2]^{1/2}+1}-1\right]&\nonumber\\
&-\frac{2}{\pi}\int_{0}^{\infty} dx \frac{x \sin(\alpha R x)}{[x^4+A^2x^2]^{1/2}+1},&\label{central}
\end{align}
where we have defined $\alpha^2=-2mE_R/\hbar^2$ and $A=1/(\alpha\xi)$. 

In the case where the bosons are non-interacting, i.e. $a_B\to 0$, we have $\xi\to \infty$ and thus $A\to 0$. 
In this limit the integrals can be performed analytically and we arrive at the well-known formula
\begin{equation}\label{eig}
\alpha R=\frac{R}{a}+e^{-\alpha R}.
\end{equation}
In the case of resonant interactions, $|a|=\infty$, the equation has the solution $\alpha R=x_0\sim 0.567$. To 
lowest order in $R/a$ (i.e. for large $|a|$ and/or small radii), we find
\begin{equation}
E_R=-\frac{\hbar^2x_{0}^{2}}{2mR^2}\left[1+\frac{1}{x_{0}^{2}(1+e^{x_0})}\frac{R}{a}\right].\label{noBEC}
\end{equation}
Another interesting limit, is $R\gg a$. In that case we can neglect the exponential term in Eq.~\eqref{eig}, and
we find a well-known result
%$\begin{equation}
$E_R=-\frac{\hbar^2}{2ma^2}.$
%\end{equation}
Notice that this {\it only} works for $a>0$, since the $a<0$ case has no solution. This energy is the 
usual energy of a particle of mass $m$ in the delta-function potential of a much heavier particle
of mass $M\gg m$ (i.e. a fixed potential center). The physical interpretation is that the small 
mass particle forms a bound state with one of the heavy particles.

\begin{figure}
\centering
\includegraphics[scale=0.45]{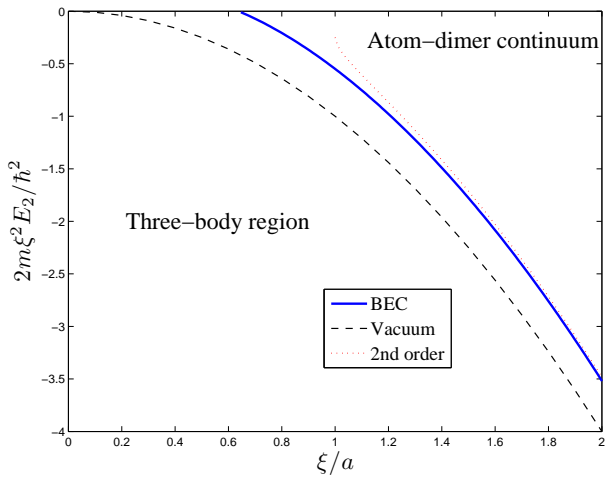}
\caption{Solution of the heavy-light two-body problem in the presence of a condensate of light particles. The
solid (blue) line shows the energy in the presence of a condensate, while the dashed (black) line is the 
corresponding vacuum solution. For $\tfrac{\xi}{a}\gg 1$, the condensate and vacuum solutions become 
indistinguishable.
The dotted (red) curve is obtained from Eq.~\eqref{tworoot}. The atom-dimer 
continuum is on the right side of the curves, while we expect universal three-body states to appear on the 
left. A clear change of threshold at small binding energy is found due to the condensate background.}
\label{twobody}
\end{figure}

\section{Heavy-light two-body states}
We first consider a single impurity. 
This problem can be handled in the 
same manner as the discussion above. The only difference is that in Eq.~\eqref{central} the 
integral with the sine term is absent. The first integral in Eq.~\eqref{central} is 
analytically tractable but the expression is long and cumbersome and does not really 
yield any insights. However, since we are concerned with the large $\xi$ limit we 
can make an expansion in $A$ inside the integral. This requires the stronger 
condition of $\alpha\xi$ large, which must be checked after doing the integral. 
After expansion one has
\begin{equation}
\int_{0}^{\infty} dx \left[\frac{x^2}{x^2+A^2/2+1}-1\right]
\end{equation}
and using this we arrive at $\frac{1}{a}=\alpha+\frac{1}{4\alpha\xi^2}$
which is a hidden second degree equation in $\alpha$ with one positive root
\begin{equation}
\alpha=\frac{1}{2a}\left[1+\sqrt{1-\frac{a^2}{\xi^2}}\right],\label{tworoot}
\end{equation}
which requires $\xi/a\geq 1$. 

In Fig.~\ref{twobody} we show the two-body binding energies of the heavy-light system with
and without the presence of a condensate. We see a clear tendency of the condensate
to push the threshold away from unitarity and into the regime of small positive $a$, i.e.
one needs stronger attraction to bind the light-heavy system in the presence of a condensate.
A shift of the two-body threshold is also found with fermionic backgrounds \cite{nygaard2011,song2011}.
However, in the light fermion and heavy impurity case the effect is opposite 
when the Fermi sea is inert (with no particle-hole excitations) \cite{nygaard2011} and pushes
the threshold to the $a<0$ region, i.e it enhances binding. Qualitatively, this 
can be understood as follows.
In the case of a Fermi gas and an impurity, the 
binding is provided by the fact that there is a Fermi surface (reducing a 
three-dimensional problem to a two-dimensional one where binding is much easier). 
Around the Fermi surface we have a linear dispersion 
of particles, similar to the situation in the present study with bosons that have
a linear dispersion around zero momentum. However, the Fermi surface moves closer 
to zero momentum as the mass of the fermions increase. This implies that 
there is a strong similarity of an impurity interacting with a Fermi sea of {\it heavy} fermions
and an impurity interacting with a condensate of {\it light} bosons.
Whether this holds as a general mapping between the two situations also
away from the extreme light/heavy mass regions
cannot be addressed within the Born-Oppenheimer approximation. This 
is an interesting question for future studies.

\section{Two impurities in a condensate}
\begin{figure}
\centering
\includegraphics[scale=0.48]{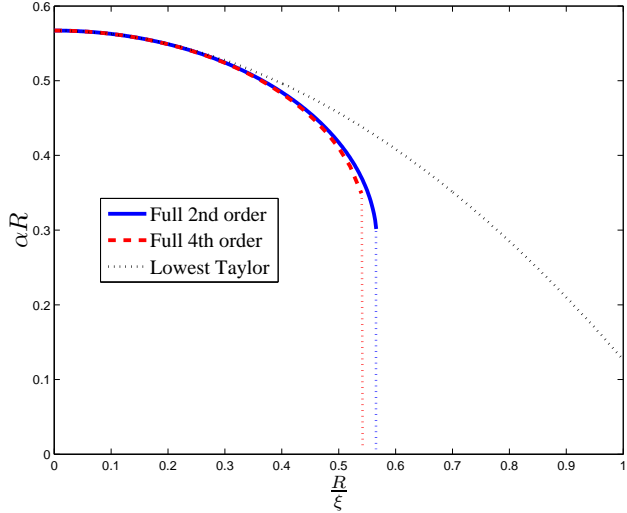}
\caption{$\alpha R$ as a function of $R/\xi$. A full solution could only be found for $R/\xi<x_0$. 
The full (blue) line is the full solution to 2nd order in $A$, while the dashed (red) shows the 
full solution to 4th order. For comparison, the dotted (black) line shows the second order expansion around zero. 
The thin dotted lines guide the eye to the positions beyond which no full solutions were found.}
\label{lambda}
\end{figure}

We now proceed to consider the effect of a condensate of light
particles on two impurities.
The solution for the potential of the two heavy impurities, 
given implicitly through $\alpha$, is found by solving Eq.~\eqref{central}.
While the first integral in Eq.~\eqref{central} is exactly solvable, 
the second term is very difficult to 
handle both analytically and numerically.
We have solved it using various 
approximation schemes and present the results in Fig.~\ref{lambda}.
The lines termed 'full' are expansion of Eq.~\eqref{central}
to 2nd and 4th order in $A$. The full solution to second
order yields the equation 
\begin{align}
\alpha R+\frac{1}{4}\alpha R A^2 = \textrm{exp}(-\alpha R-\frac{1}{4}\alpha R A^2),\label{eigeq}
\end{align}
from which the 'lowest Taylor' is obtained by second order 
expansion around $R/\xi=0$.
In the absence of the condensate, the solution is given by the
first term in Eq.~\eqref{noBEC} which can be written $\alpha R=x_0$.
We thus see that the effect of the condensate and the corresponding 
change in the dispersion of the light particle, is to suppress the 
attraction between the impurities. The scale of this suppression 
is not surprisingly $\xi$. We arrive at the conclusion that 
the presence of a condensate of light particles will tend to 
make universal three-body states more difficult to form in this
non-trivial many-body background.

Numerically, it becomes increasingly difficult to find a solution as $R/\xi\to x_0$, and beyond this
point we have failed to find solutions. This indicates that the expansion of the integrand before
performing the integrals becomes problematic.
Fig.~\ref{lambda} does indicate that the quantity $\alpha R$ goes to zero at some finite value of $R/\xi$. 
Numerically, we find that a good fit to the solutions for $\alpha R>x_0/2$ is 
\begin{align}
\alpha R =x_0\left[1-\left(\frac{R}{x_0\xi}\right)^2\right]^{1/2}.
\end{align}
Another way to approach the limit where $\alpha\xi$ becomes small 
is to consider the case $A^2\gg 1$ and make this 
approximation before doing the integrals in Eq.~\ref{central}. 
The first integral is analytic and can be easily expanded, while the second integral
can be handled by using $\sqrt{x^4+A^2x^2}\to Ax$ for $A^2\gg 1$, taking a sine transform
and subsequently doing the expansion. This yields the expression
\begin{align}
(\alpha\xi)^2=\frac{2}{\pi}\frac{\frac{\xi}{R}-\frac{R}{\xi}}{\frac{R}{\xi}-1}<0,
\end{align}
which is always negative. This indicates that $E_R$ does indeed change sign as $R\sim \xi$. However,
taking the limit of $R\to\xi$ in this result yields $-\tfrac{4}{\pi}$ and $A^2$ is now of order one
which is in conflict with the limit we used to derive the expression. However, it strongly suggests 
that $E_R$ will go to zero for some value of $R/\xi$.

\begin{figure}
\centering
\includegraphics[scale=0.48]{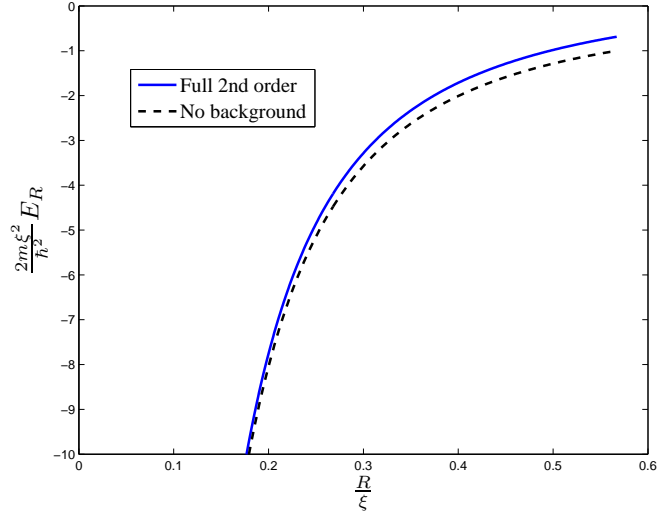}
\caption{Effective potential for the two heavy impurities, $E_R$, as a function of their relative distance
$R$. The solid (blue) curve shows the result from a second order expansion of Eq.~\eqref{central} with 
respect to $(\alpha\xi)^{-1}$, while the dashed (black) curve gives the results with no condensate
background ($-\tfrac{x_{0}^{2}}{R^2}$). The end point of the solid curve is at $R/\xi=x_0$ where the
second order solution ceases to exist. However, our results generally indicate that the potential $E_L$ 
goes to zero {\it faster} than $R^{-2}$ for large $R/\xi$.}
\label{pot}
\end{figure}

Our results for the potential between the two heavy impurities, $E_R$, are plotted in Fig.~\ref{pot}.
We see that the presence of the condensate tends to make the potential go to zero faster than $R^{-2}$
as $R$ increases. An effective repulsive effect is thus seen to originate from the many-body
background in the light particle component. The Efimov 
effect depends on the $R^{-2}$ functional form of the potential, and of course on the fact
that it is attractive \cite{jensen2004,braaten2006}. Our results imply that the condensate coherence length, $\xi$, 
must be considered when estimating the Efimov effect in a degenerate Bose gas setting. 
Moreover, the length scale of the modification is $\xi$ to within a factor of order one. 
We therefore conclude that at unitarity, the number of universal three-body 
bound state can be estimated by 
analogy with Efimov's original formula \cite{efimov1970} and becomes
\begin{equation}
N_B\approx\frac{s_0}{\pi}\textrm{log}\left(\frac{\xi}{R_0}\right),\label{number}
\end{equation}
where $s_0$ is the scale factor and $R_0$ is a short-distance cut-off \cite{efimov1970}.
More generally, we expect that whichever is smaller of $|a|$ and $\xi$ will cut off the number of states
allowed in the spectrum.

We note that our findings do not indicate any modification of the
potential for small distances. This can be understood from the dispersion relation 
of the light particles which become quadratic at high momenta corresponding to 
short distance. Condensates modify behavior at long-range (where low-energy universal three-body states
may reside), short distance is left
more or less undisturbed. This means that the short distance cut-off that is necessary
to bound the attractive $R^{-2}$ and avoid collapse is provided by short-distance 
physics ($R_0$ in Eq.~\eqref{number}, often called the three-body parameter). 
However, recent studies have found a 
strong connection between $R_0$ and the two-body physics of the systems as given
by the van der Waals length, 
$r_\textrm{vdW}$ \cite{berninger2011,naidon2011,chin2011,schmidt2012,wang2012,peder2012,naidon2012}. 
In the case studied here where the impurity is much heavier than the light constituent, 
$R_0$ is given simply by the details of the heavy-heavy system as is seen from the 
derivation of $E_R$ above (and from the numerical results of Ref.~\cite{wangwang2012}).
We therefore expect that $R_0\sim r_\textrm{vdW}$ to within factors of order unity 
(between $\sqrt{2}$ \cite{peder2012} and 2 \cite{wang2012}). Note that 
we are assuming that there are no resonances in the heavy-heavy system. This
should always be closely investigated in realistic setups where interactions are
typically controlled by external fields \cite{chin2010} that affect both
heavy-light and heavy-heavy systems.

\section{Experimental Considerations}
We now discuss some systems of experimental relavance to the physics studied above. 
Light alkali atoms that have been condensed are $^{7}$Li and $^{23}$Na, although
the latter is not really light when taking ratios with other interesting systems. 
However, recent advances in studies of metastable Helium, $^{4}$He$^*$ \cite{borbely2012,knoop2012a} and the
potential of those experiments to make mixtures of $^{4}$He$^*$ and $^{87}$Rb makes
that a very interesting system \cite{knoop2012b}. Likewise, the mixture of 
$^7$Li and $^{133}$Cs \cite{mudrich2002} or $^7$Li and $^{87}$Rb \cite{marzok2009}
would be favourable, and also potential mixtures of Ytterbium isotopes (mass numbers
168-176) \cite{onomoto2007,kitagawa2008} and $^7$Li or $^4$He.

The results we have presented here show that the Efimov spectrum can be 
modified at large distances by the presence of a condensate background, and also
that there is no effect at short distance from the condensate. If we insert
the definition of $\xi$ into Eq.~\eqref{number}, we find the following 
formula for the expected number of three-body Efimov states at unitarity 
(which effectively assumes that $|a|\gg\xi$) for our setup
\begin{align}
N_B=\frac{s_0}{2\pi}\,\textrm{ln}\left[2.7\cdot 10^{10}\frac{a_0}{a_B}\left(\frac{a_0}{r_\textrm{vdW}}\right)^2\frac{10^{13}\,\textrm{cm}^{-3}}{n_0}\right],
\end{align}
where $n_0$ is the condensate density, $a_B$ is the scattering length of the condensed bosons, and 
$r_\textrm{vdW}$ is the two-body van der Waals length associated with the interatomic potential 
of the two heavy impurities. The unit $a_0$ is the Bohr radius. 
If we consider the $^4$He-$^{87}$Rb or $^7$Li-$^{133}$Cs cases, 
the mass ratios are roughly the same and we have $s_0=1.98$. For the $^4$He$^*$-$^{87}$Rb case, 
$a_B\sim 142a_0$ \cite{borbely2012,knoop2012b} and $r_\textrm{vdw}(^{87}\textrm{Rb})\sim 83a_0$ \cite{chin2010}, 
and we obtain $N_B\sim 3.2$ at $n_0=10^{13}\,\textrm{cm}^{-3}$. 
One order of magnitude increase in $n_0$ or $a_B$
brings this down to $N_B\sim 2.5$, so we see a sizable effect. Of course we 
are assuming that there are non-overlapping resonances in $^{4}$He-$^{87}$Rb
and $^{87}$Rb-$^{87}$Rb which is currently unknown. The example of 
$^7$Li-$^{133}$Cs is slightly more complicated since 
$^{7}$Li has attractive interactions at zero magnetic fields \cite{chin2010}. 
However, let us for the moment assume that we can tune the scattering length 
away from the attractive region and also find a good resonance in the 
Li-Cs system (as recently done for the closely related
case $^{6}$Li-$^{133}$Cs \cite{repp2012,tung2012}). If we assume that 
$a_B\sim 100a_0$ and use $r_\textrm{vdW}\sim 101a_0$ \cite{chin2010}, 
we find more or less exactly the same scenario as in the $^4$He-$^{87}$Rb
case. We thus see that the effect of condensation of light particles
in heavy-heavy-light three-body systems should be accessible in current 
experiments.
 
Other mixtures are being pursued at the moment that are relevant for our purposes
since they contain Bose components that can be condensed, but for which the 
mass ratios are too large for the Born-Oppenheimer approximation to be fully
justified. Examples are $^{23}$Na mixed with $^{40}$K \cite{wu2012} and $^{87}$Rb
with $^{133}$Cs \cite{takekoshi2012}. For the latter system the experiments
that are probably most interesting for our purposes are those with a few heavy impurities
in a condensate of the lighter species by the Widera group \cite{weber2010,spethmann2011}.
While the approximations used here are not expected to be accurate in relation to 
these less mass imbalanced mixtures, we do expect that similar signatures should 
occur, and that the Efimov effect, if present, will be modified by a condensate
background when $\xi$ is sufficiently small.

\section{Discussion}
We have considered the influence of a condensate background on three-body bound state
physics in the case of two heavy impurity atoms embedded in a condensate of light
particles and assuming that the light-heavy interaction is short-ranged. Using 
the Born-Oppenheimer approximation we calculate the modification of the 
heavy-heavy interatomic potential when the light particle dynamics is 
integrated out. Our results demonstrate that this potential is strongly
modified at length scale corresponding to the condensate coherence length,
and eventually turns from attractive inverse square (necessary for 
the Efimov effect), through zero, and then most likely into a repulsive
potential at very large distance. These findings indicate that the 
coherence length must be considered as a length scale when estimating the
potential for such systems to form universal three-body bound states. 
In the case where the heavy-light interaction is resonant (infinite
scattering length), the coherence length replaces the scattering 
length in the famous Efimov formula, Eq.~\eqref{number}. We have 
estimated the effects of our findings on experimental mixtures of 
current interest and find that by tuning interaction strengths and 
condensate density, it should be possible to manipulate the number 
of universal three-body states.

In future studies it is necessary to go beyond the Born-Oppenheimer
approximation in order not to rely on large mass imbalanc. Furthermore,
two or three of the constituents may be condensed and it would interesting
to study the effects that this will have on the three-body spectrum. 
From the current study we would expect that some combination of 
coherence lengths and scattering lengths would decide the number of 
bound states. It would also be interesting to study the addition of 
degenerate Fermi components. Here the coherence length is replaced
by the Fermi momentum \cite{macneill2011,nygaard2011}.

Another important way in which to improve the current study is 
in the description of the Bose condensate through the Bogoliubov
formalism. If we have a strongly interacting condensate one needs
to consider modifications here. Also the neglect of states with 
quasi-particles (phonons) may not be justified then and we need
to consider instead a starting point in line with the Fr{\"o}hlich
polaron system \cite{devreese2009}. However, we do expect to 
see qualitatively the same physics, i.e. that presence of 
backgrounds will modify the universal three-body spectrum and 
in many cases suppress bound state formation.

The essential assumption used here was that the light particle 
had a linear dispersion at low energy. This is reminiscent of the 
low-energy dispersion of a system like Graphene \cite{neto2009}
or the surface of a topological insulator \cite{hasan2010}
where it is electrons that have linear dispersion at low-energy.
Adding impurities to such systems and studying bound states
induced by the electronic surroundings with linear dispersion is 
an interesting prospect.

This work was supported by the Danish 
Agency for Science, Technology, and Innovation under the 
Danish Council for Independent Research - Natural Sciences. We
are grateful to Manuel Valiente for feedback on the manuscript.

\end{document}